# Extending the observation limits of Imaging Air Cherenkov Telescopes toward horizon

Razmik Mirzoyan[a,*], Ievgen Vovk[a,*], Michele Peresano[b], Petar Temnikov[c], Darko Zaric[d], Nikola Godinovic[d],

Juliane van Scherpenberg[a], Juergen Besenrieder[a], Masahiro Teshima[a]

On Behalf of the MAGIC Very Large Zenith Angle Observation Working Group

[a] *Max-Planck-Institute for Physics, Munich, Germany*

[b] *Università di Udine and INFN, sezione di Trieste, Italy, Udine, Italy*

[c] *Institute for Nuclear Research and Nuclear Energy, Sofia, Bulgaria*

[d] *Croatian MAGIC Consortium: University of Split, Croatia*

**Abstract**

Usually the Imaging Atmospheric Cherenkov Telescopes, used for the ground-based gamma-ray astronomy in the very high energy range 50 GeV - 50 TeV, perform air shower observations till the zenith angle of ~60°. Beyond that limit the column density of air increases rapidly and the Cherenkov light absorption starts playing a major role. Absence of a proper calibration method of light transmission restrained researchers performing regular measurements under zenith angles >>60°. We extend the observation of air showers in Cherenkov light till almost the horizon. We use an aperture photometry technique for calibrating the Cherenkov light transmission in atmosphere during observations under very large zenith angles. Along with longer in time observations of a given source, this observation technique allows one to strongly increase the collection area and the event statistics of Cherenkov telescopes for the very high energy part of the spectrum. Study of the spectra of the highest energy gamma rays from a handful of candidate sources can provide a clue for the origin of the galactic cosmic rays. We show that MAGIC very large zenith angle observations yield a collection area in excess of a square kilometer. For selected sources this is becoming comparable with the target collection area anticipated with the Cherenkov Telescope Array.

PeVatron, very large zenith angle observations

## 1. Introduction

Observation of very high energy cosmic and gamma rays are associated with low event count rates because their fluxes are inversely proportional to energy, i.e. the higher the energy the less is the number of events. Collection area $A_{eff}$ of extended air showers (EAS) by an Imaging Atmospheric Cherenkov Telescopes (IACTs) is determined by the size of the Cherenkov light pool on the observation level. For close to vertical (zenith) observations the collection area of a single telescope does not significantly exceed ~0.05 km$^2$ on the observation height of ~2km a.s.l.. This collection area can be significantly increased by using an array of similar, largely separated telescopes. This is the approach followed by the forthcoming Cherenkov Telescope Array (CTA) collaboration [1]. Alternatively increase in the collection area can be achieved by performing observations at large zenith angles [2, 3]. During such observations the column density of air in a given direction is significantly larger compared to that from the vertical direction, see Fig.1. For example, an air shower of 1TeV energy reaches the maximum of its development after traversing ~300 g/cm², counted from the top of the atmosphere. Obviously this is independent on the observation angle (the entire atmosphere is ~1036 g/cm² in the vertical direction and it is ~6 times "thicker" during observations at the zenith angle of ~80°). An air shower from large zenith angle will develop far away from the observing telescope. At small zenith angle observations the Cherenkov light from EAS will propagate through a relatively short distance. In contrast, during very large zenith angle (VLZA; > 70°) observations the light will propagate through much thicker atmosphere and longer distance.

---

[*] Corresponding authors: R. Mirzoyan (Razmik.Mirzoyan@mpp.mpg.de); I. Vovk (Ievgen.Vovk@mpp.mpg.de)

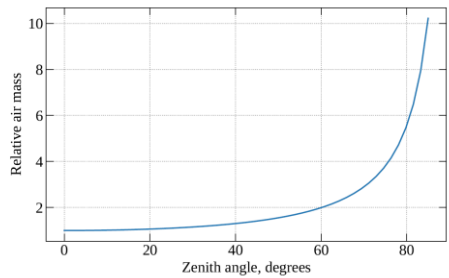

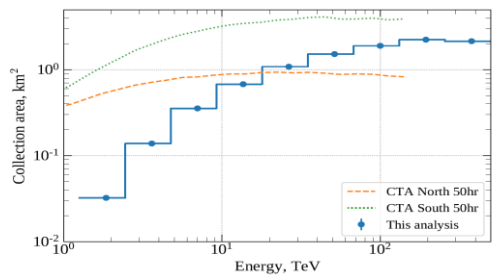

Fig.1. Relative air mass versus zenith angle.

Fig.2. Collection area of MAGIC for VLZA observations in the range (70-80)°. The collection area of CTA for both the Northern and the Southern locations are shown for comparison.

The range of the Cherenkov light emission angle is similar for both the zenith and the VLZA observations. Due to the above mentioned two reasons a given energy shower observed under VLZA will illuminate a much larger area on the observation level (albeit at the much lower density of photons) than when observed under low zenith angles. At VLZA observations the very high energy showers, which produce lot of photons, can be detected at large impact distances (~1km) from the telescope, thus providing high event statistics.

Study of the spectra of the highest energy gamma rays from a number of selected sources, known as the PeVatron candidates, can provide a clue for the origin of the galactic cosmic rays.

We report about the VLZA observation technique used with the MAGIC telescopes, which allow us observing air showers almost till the horizon. This technique significantly increases the collection area and event statistics of air showers for energies above 10 TeV. To assure the proper calibration of the observational data, we developed an aperture photometry technique for measuring the (contemporary with gamma-ray observations) atmospheric transmission. These latter employ optical images of the stellar field next to the source position and provide accuracy better than 10 %.

Below we present the technique and details of observation at VLZA with the MAGIC telescopes.

### 1.1. The MAGIC Telescopes

MAGIC (Major Atmospheric Gamma Imaging Cherenkov) consists of two 17 m diameter IACTs, separated by 85 m distance and located at an altitude of 2200 m a.s.l. at the *Roque de los Muchachos* European Northern Observatory on the Canary island of La Palma, Spain (28°45' N, 17°53' W).

The telescopes are used to image flashes of Cherenkov light produced by Extensive Air Showers (EAS) initiated in the upper atmosphere by hadrons, electrons and gamma-ray photons with energies $\geq$ 30 GeV. The telescopes are operated in hardware coincidence (so-called stereo) mode. For E $\geq$ 220 GeV and low zenith angle ($\leq 30°$) observations the integral sensitivity of MAGIC is (0.66 ± 0.03) % in units of the Crab Nebula flux (C.U.) for 50 hours of observations [4].

## 2. MAGIC very large zenith angle observations

*2.1. Light Arriving to the Telescope*

By using a handheld mini-spectrometer (type Hamamatsu C10082CAH) we performed observations of the spectrum of the sun on a clear day at the *Roque de los Muchachos* observatory. On Fig.3 one can see overlaid spectra of the sun measured in the zenith angle range of (65.4 – 89.9)°. The lowest curve is measured at 89.9°, i.e. almost from the horizon, which is ~170 km far from the observer. The lowest curve shows the spectrum range (~500 – 650) nm where light can be detected by using, for example, classical bialkali photo multiplier tubes (PMT). This shows us the range of the Cherenkov light spectrum, which from a close to horizon remote air shower can reach the observation level.

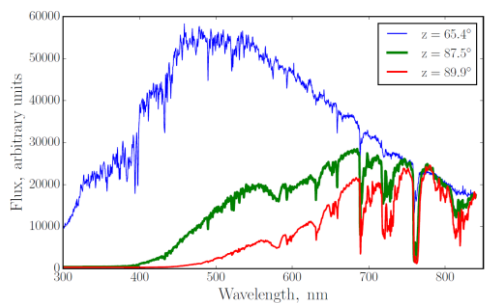

Fig.3. Spectra of the sun measured in the zenith angle range (65.4 – 89.9)°. The area under the bottom curve shows the range of the spectrum arriving to the observer almost from the horizon, from the distance of ~170 km.

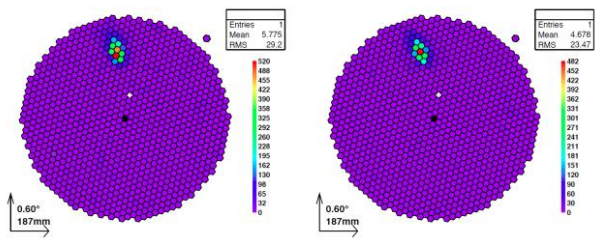

Fig.4. Example of a gamma-ray candidate event observed by MAGICs under the zenith angle of 77.8°; the estimated energy is 144.4 TeV. The white cross denotes the position of the observed source.

*2.2. Importance of the High Resolution Imaging Camera for VLZA Observations*

Major part of light from an EAS is emitted by its maximum development region. At the VLZA observations the shower maximum is far from the telescope. Because of this the size of the observed image size shrinks and one needs to provide fine resolution camera for triggering and measuring the image of a remote air shower. MAGIC is using 1039 PMT-based pixels of aperture 0.10° in its imaging cameras, set to the condition of three next-neighbour pixel trigger from each telescope for measuring a coincidence event. When observing under ~80° zenith angle, although the majority of images are produced by multiple tens of TeV energy air showers, their geometrical size is becoming small. In Fig.4 we show example of an image of a gamma-candidate event, observed from a source under the VLZA of 77.8°; its estimated energy is 144.4 TeV.

*2.3. VLZA Observations*

VLZA observations can be performed when the source is rising above the horizon and when it is setting. These provide somewhat different sensitivity for MAGIC to the impinging flux of γ-rays due to the varying projected distance between the telescopes seen from the direction of the source.

During TeV observations performed under VLZA the distance to shower maximum is on the order of ~(50-100) km from the telescope (as opposed to < 10 km at lower zenith angles ≤ 30°). Obviously VLZA measurements are subject to strong light attenuation due to the scattering and absorption in the atmosphere. At low zenith angles MAGIC is using a micro-LIDAR [5], which allows us to probe the atmospheric absorption at distances ≤ 20 km. This is not sufficient for VLZA observations. For monitoring and calibrating the atmospheric attenuation during the VLZA observations we are continuously integrating images of the stellar field next to the observed target. For this purpose we are using identical CCD cameras (type SBIG STL 1001E coupled to AF Nikkor 180mm f/2.8 lens from Nikon) on the telescopes, which are set close to the center of the mirror dish. A remotely controlled rotating filter wheels with red, green, blue and luminance filters are set in front of the CCD cameras. This setup allows us to monitor the transmission of atmosphere in the chosen direction with accuracy better than 10 %. Incidentally, this same setup allows us to also measure the absolute reflectance of the mirror dishes of both MAGIC-I and MAGIC-II telescopes [6].

During the VLZA data taking the energy threshold of MAGICs quickly increases from ~1 TeV at zenith angle of 70° to ~10 TeV when approaching 80°. The collection area for energies above 10 TeV increases by more than one order of magnitude, reaching ~1 km². This allows us probing the spectrum of a selected source candidate at the highest energies.

To estimate the collection area for the VLZA observations under (70-80)°, we performed dedicated Monte Carlo (MC) simulations. These were done by using the Corsika v6.99, which included the curvature of the atmosphere.

The resulting collection area estimated after the data selection cuts is shown in Fig.2. For comparison, also the expected collection area of the currently under construction CTA[2] array in both Northern and Southern locations are shown.

*2.4* Atmospheric Transmission Under VLZA Observations

Atmospheric transmission defines the amount of Cherenkov light that can reach a telescope. The uncertainty in it contributes to the overall uncertainty of the energy scale of measured events.

To estimate the atmosphere transmission, images of the stellar field next to a selected source were taken with CCD every 90 seconds, periodically changing the colour filters from red ($\lambda_{mean}$ ~ 640 nm) to green ($\lambda_{mean}$ ~ 530 nm) and to blue ($\lambda_{mean}$ ~ 450 nm). The acquired images were flat-fielded and cleaned from hot pixels and dark current contribution. Then counts from selected bright stars in the close vicinity of the observed target were estimated as a difference of counts from the circular region around the star and the background counts from an annular region of somewhat larger diameter.

In order to calibrate this aperture photometry procedure, an additional imaging of this stellar field was performed on several very clean nights. These were selected by the high and stable LIDAR transmission at low zenith angle observations, at the absence of clouds and law amount of dust in the air. During such nights light absorption follows the Lambert-Beer's law:

$$c = c_0 exp(-\alpha m_{air}(z)) \qquad (1)$$

where $c$ is the number of background-subtracted CCD counts, $c_0$ is the number of counts before absorption, $\alpha \approx const$ is the specific absorption coefficient and $m_{air}$ is the air mass at a given zenith angle $z$.

The constant $c_0$ can be determined from *Eq.(1)* by using the measured CCD counts from the selected star observed under different zenith angles. Knowing $c_0$, the average absorption coefficient $\alpha$ during the subsequent observational sessions can be estimated as $\alpha = -log(c/c_0)/m_{air}(z)$.

Contemporaneous observation of selected stars during the VLZA data taking allowed us estimating atmospheric transmission for EAS with temporal resolution of 1.5-3 minutes. The maximum height of every shower, estimated as a part of the standard data analysis software package, is used to compute the line of sight distance to the shower maximum and to derive the corresponding value of the air mass $m_{air}^{EAS}$. The resulting absorption can then be estimated as $\tau_{data} = exp(-\alpha m_{air}^{EAS}(z))$. The ratio of this latter value to the absorption assumed in the MAGIC detector MC simulations (for the same zenith angle and shower distance) then defines the relative light scale $s = \tau_{data}/\tau_{MC}$, which is finally used for correcting the estimated event energies; for details see [7].

Though the stellar light measurements, described above, provide a simple and reliable way to estimate the total atmospheric transmission, they are subject to inaccuracies due to the uncertainties in the derived calibration constants $c_0$ and uncertainties in the measured CCD counts during the observations. We have minimized the latter by choosing the camera exposure time to integrate ~ 30k CCD counts from the selected stars. Under that condition the resulting uncertainty stays below 1 %.

---

[2] Expected CTA performance can be found under: *https://www.cta-observatory.org/science/cta-performance*

The uncertainty on the calibration constants $c_0$ was computed from several $c_0´$ estimates, taken on nights with stable transparency of atmosphere. The standard deviation of these estimates suggests that the calibration constants for the reference stars are determined with the accuracy ≤ 5 %.

Currently we are developing a second method for measuring the atmospheric absorption during observations of sources under VLZA. It is based on the use of a small telescope of an aperture of 28cm, coupled to a spectrograph. This telescope will track a relatively bright star in the close proximity of the selected source. The periodically acquired spectra from such a star will allow us to measure the momentary transmission of the atmosphere with a better precision than with the above described CCD-based method.

*2.5 MC-Data Comparison*

EAS development observed under zenith angles above 70° proceeds primarily in the rarefied layers of the upper atmosphere at ~(50-100) km (or even further) distance from the observer. MC simulations reveal certain peculiarities in the shower evolution, which depend on the origin of the primary particle; for details see [8].

The technique of VLZA observations is a new terrain. Compared to lower zenith angles one may anticipate larger MC to data discrepancy, especially at observations performed under close to horizon extremely large zenith angles.

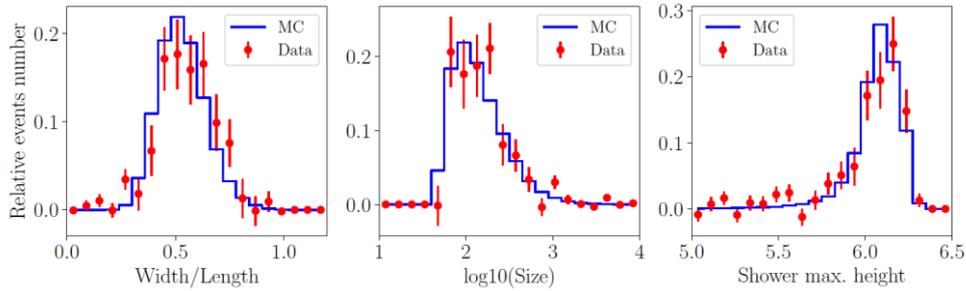

Fig.5. Comparison of the main parameters of the MC simulated (blue) and real (red) event data sets, recorded in the (70-75)° zenith angle range. *Size*, *Length* and *Width* are the so-called *Hillas* parameters [9], whereas the shower maximum height is reconstructed from the MAGIC data by using the standard analysis pipeline.

We have compared the distribution of the basic EAS "Hillas" parameters *Size*, *Length*, *Width* [9] as well as the shower maximum height in MAGIC MC data set with the excess distributions of the same parameters in the on- and off-source regions for a selected source. These were derived with loose event selection cuts. This comparison is shown in Fig.5 for MC and real events observed in the zenith angle range (70-75)°.

As one can see from this figure, no significant difference is present between the real observed and simulated by MC VLZA data.

The detailed performance of the MAGIC telescope observations of a selected source under the VLZA will be presented in a forthcoming paper.

**3. Conclusions**

We have developed a VLZA observation technique, which provides significant increase in the collection area of air showers for the > 10 TeV part of the spectrum. This technique enables reaching a collection area in excess of km², which can allow one studying the spectrum of highest energy gamma-rays emitted by the PeVatron candidate sources.

**Acknowledgments**

We want to thank our colleagues from the MAGIC collaboration for the continuous interest and support.